\documentstyle[epsfig]{elsart}

\begin{document}

\begin{frontmatter}

\title{Crossover to Non-universal Microscopic Spectral Fluctuations in
  Lattice Gauge Theory}

\author[KL]{M.E.~Berbenni-Bitsch},
\author[R]{M.~G\"ockeler},
\author[HD]{T.~Guhr},
\author[CO]{A.D.~Jackson},
\author[HD]{J.-Z.~Ma},
\author[KL]{S.~Meyer},
\author[R]{A.~Sch\"afer},
\author[HD]{H.A.~Weidenm\"uller},
\author[M]{T.~Wettig}, and
\author[HD]{T.~Wilke}

\address[KL]{Fachbereich Physik -- Theoretische Physik, Universit\"at
  Kaiserslautern, D-67663 Kaiserslautern, Germany}
\address[R]{Institut f\"ur Theoretische Physik, Universit\"at
  Regensburg, D-93040 Regensburg, Germany} 
\address[HD]{Max-Planck-Institut f\"ur Kernphysik, Postfach 103980,
  D-69029 Heidelberg, Germany}
\address[CO]{Niels-Bohr-Institute, Blegdamsvej 17, DK-2100 Copenhagen
  \O, Denmark} 
\address[M]{Institut f\"ur Theoretische Physik, Technische
  Universit\"at M\"unchen, D-85747 Garching, Germany}

\begin{abstract}
  The spectrum of the Dirac operator near zero virtuality obtained
  in lattice gauge simulations is known to be universally described
  by chiral random matrix theory. We address the question of the 
  maximum energy for which this universality persists. For this purpose, 
  we analyze large ensembles of complete spectra of the Euclidean Dirac 
  operator for staggered fermions. We calculate the disconnected scalar
  susceptibility and the microscopic number variance for the chiral
  symplectic ensemble of random matrices and compare the results with
  lattice Dirac spectra for quenched SU(2). The crossover to a
  non--universal regime is clearly identified and found to scale with
  the square of the linear lattice size and with $f_{\pi}^2$, in
  agreement with theoretical expectations.
\end{abstract}

\date{25 June 1998}

\end{frontmatter}

\maketitle

Recently, it has been shown by several authors that chiral random
matrix theory (chRMT) is able to reproduce quantitatively spectral
properties of the Dirac operator obtained from QCD lattice data. This
statement is valid both for fluctuation properties in the bulk of the
spectrum and for microscopic spectral properties near zero virtuality,
see the reviews~\cite{r1,r2} and Refs.\,\cite{r3,r4,r5,Ma}.  This
result implies that the spectral fluctuation properties of the Dirac
operator are universal, i.e., determined solely by the underlying
symmetry of the problem and quite independent of specific aspects of
QCD.  The success of chRMT poses the question: Which QCD energy scale
limits this universal behavior? In mesoscopic physics, the analogous
scale (i.e., the ``Thouless energy'') is given by $E_C \sim L^{-2}$,
where $L$ is the length of the sample.  Spectral fluctuation
properties of a mesoscopic probe obey random matrix theory only in
energy intervals smaller than $E_C$.

Two recent publications \cite{Zahed98,Jac98} address the existence of
such a scale, here denoted by $\lambda_{\rm RMT}$, in QCD.  (Earlier
qualitative discussions of the transport properties of light quarks in
the QCD vacuum can be found in Refs.~\cite{Bank80,Diak}, and a more
quantitative approach was taken recently in Ref.~\cite{Stern}.)  The
scale $\lambda_{\rm RMT}$ is important since on smaller scales, QCD
calculations do not contain system--specific information.  The authors
of Ref.~\cite{Zahed98} used general arguments and simple estimates for
$\lambda_{\rm RMT}$, while Ref.~\cite{Jac98} provides
semi--quantitative results for $\lambda_{\rm RMT}$ based on the
instanton liquid model. It is the purpose of this Letter to deduce for
the first time values for $\lambda_{\rm RMT}$ directly from
microscopic QCD lattice data and to establish the scaling properties
of this quantity both with respect to lattice size and coupling
constant. A recent analysis of spectral data in the bulk \cite{gu1}
yields results which are consistent with our findings.

We recall that chRMT uses a generating functional of the form
\begin{equation}
\label{func}
Z^{\beta_{\rm D}}_{N_f}=\int D[W] \prod_{f=1}^{N_f} {\rm det}({\cal D}+m_f)
e^{-{N\beta_{\rm D}\over 4} \,{\rm tr}\, v(W^{\dagger}W)}
\end{equation}
with
\begin{equation}
  \label{D}
{\cal D}= \left( \begin{array}{cc}
0 & iW\\
iW^{\dagger} &0 
\end{array} \right)  
\end{equation}
and a potential, $v$, which determines the distribution of the matrix
elements of $W$. The universal spectral fluctuation properties do not
depend on the choice of $v$ \cite{Ak1} which is taken to be a Gaussian 
for convenience,
\begin{equation}
v(W^{\dagger}W)= \Sigma^2 W^{\dagger}W\:.
\label{e5}
\end{equation}
In Eq.\,(\ref{func}), we consider only the sector of topological
charge zero because our lattice data agree with the chRMT results in
this sector \cite{r5}.  With $N$ the dimension of the matrix in
Eq.\,(\ref{D}), $\Sigma$ is the absolute value of the chiral
condensate, $\langle\bar\psi\psi\rangle$ (per flavor).  Various gauge
theories have different symmetries and, hence, different values for
$\beta_{\rm D}$.  (The index D for Dyson serves to distinguish the
symmetry parameter from the square of the inverse coupling constant
denoted by $\beta$.)  For SU($N_c$) and $N_c\geq 3$ one has
$\beta_{\rm D} = 2$ (chiral Gaussian Unitary Ensemble, chGUE); for
$N_c=2$ and staggered fermions (this is our case) one has $\beta_{\rm
  D} = 4$ (chiral Gaussian Symplectic Ensemble, chGSE); and for
$N_c=2$ and fermions in the fundamental representation one has
$\beta_{\rm D} = 1$ (chiral Gaussian Orthogonal Ensemble, chGOE), see
Ref.~\cite{Jac1}.  When we apply chRMT to quenched lattice
calculations, the determinant in Eq.\,(\ref{func}) is absent.

Earlier comparisons have shown that all predictions of chRMT such as
sum rules, microscopic spectral distributions, spectral correlations
in the bulk, nearest--neighbor spacing distributions, etc.\ agree very
well with lattice data. The single parameter of the model, $\Sigma$,
can be determined from the lattice data \cite{r5} via the Banks-Casher
relation \cite{Bank80}. Then, the chRMT predictions are parameter
free.

In Refs.\,\cite{Zahed98,Jac98}, it was argued that $\lambda_{\rm
  RMT}$ can be estimated with the help of the Gell-Mann--Oakes--Renner
relation, which yields
\begin{equation}
\label{lambda}
  \lambda_{\rm RMT} \sim {f_{\pi}^2\over \Sigma V^{1/2}}\ ,
\end{equation}
where $f_\pi=93$ MeV is the pion decay constant and $V=L^4$ is the
space-time volume.  On the lattice, $V=Na^4$, where $N$ is the number
of lattice sites and $a$ is the lattice constant which we set to unity
unless otherwise indicated.  Using the mean level spacing at zero,
$\Delta=\pi/(\Sigma V)$, Eq.\,(\ref{lambda}) can be expressed in
dimensionless form,
\begin{equation}
\label{u}
\lambda_{\rm RMT}/\Delta  \sim \frac1\pi f_{\pi}^2 L^2 \ .
\end{equation}

To determine $\lambda_{\rm RMT}$ and to test the expected dependence
of $\lambda_{\rm RMT}$ on $L$ and $\Sigma$, we use the disconnected
spectral susceptibility $\chi^{\rm disc}$ and the $\Sigma^2(0,S)$
statistic. In order to avoid confusion between the latter quantity and
the value, $\Sigma$, of the chiral condensate, we will display the
arguments $(0,S)$ in $\Sigma^2(0,S)$. We denote the limiting scale for
chRMT determined from $\chi^{\rm disc}$ by $\lambda_{\rm RMT}$ and
that determined from $\Sigma^2(0,S)$ by $S_{\rm RMT}$.  We shall see
that $\lambda_{\rm RMT}/\Delta$ and $S_{\rm RMT}$ agree within the
accuracy of our analysis, although the errors associated with $S_{\rm
  RMT}$ are larger than those for $\lambda_{\rm RMT}/\Delta$.  We
shall not address the question of how these quantities are related to
an intrinsically defined energy scale, cf.\ Ref.\,\cite{r9}.

The disconnected spectral susceptibility, $\chi^{\rm disc}$, is
defined in terms of the Dirac eigenvalues, $\lambda_k$, obtained in
lattice simulations by
\begin{equation}
 \chi^{\rm disc}=\frac{1}{N}\left\langle\sum_{k,l=1}^N
    \frac{1}{(i\lambda_k+m)(i\lambda_l+m)}\right\rangle - \frac{1}{N}
  \left\langle\sum_{k=1}^N\frac{1}{i\lambda_k+m}\right\rangle^2,
\label{e4}
\end{equation}
where the average is over independent gauge field configurations and
where $m$ denotes the valence quark mass.  Note that for SU(2) all
eigenvalues are two--fold degenerate; for, e.g., SU(3) the sums would
run up to $3N$. We study $\chi^{\rm disc}$ at zero temperature.
Lattice QCD studies of the disconnected and connected susceptibilities
at finite temperature do exist \cite{Karsch}. In this case, chRMT must
be supplemented by non--random terms which are model--dependent
\cite{r10}.  Nevertheless, the universality of the random--matrix
results is expected to persist for energies below $\lambda_{\rm RMT}$
if the (model--dependent) temperature--dependence of $\Sigma$ is taken
into account \cite{JSV}.  However, we shall not address the question
of finite temperature in this work.

The sums in Eq.\,(\ref{e4}) can be written as integrals involving the
microscopic spectral densities of the Dirac operator, i.e., the
spectral densities on the scale of the mean level spacing near zero.
We have
\begin{equation}
\chi^{\rm disc}=4u^2 \left[ \int_0^{\infty} dx {\rho_1(x)\over
    (x^2+u^2)^2}
-\int_0^{\infty} dx\int_0^{\infty} dy {\tau_2(x,y)\over
    (x^2+u^2)(y^2+u^2)}\right],
\label{e2}
\end{equation}
where $u=mN\Sigma$ and $\chi^{\rm disc}$ has been rescaled by
$1/(N\Sigma^2)$ so that all quantities in Eq.\,(\ref{e2}) are
dimensionless.  The function $\tau_2(x,y)$ is the connected part of
the microscopic spectral two-point function,
$\rho_2(x,y)=\rho_1(x)\rho_1(y)-\tau_2(x,y)$.  Equation\,(\ref{e2}) is
universal in the sense that all reference to the parameter $\Sigma$,
which depends on the simulation parameter $\beta=4/g^2$, has been
eliminated.  We now make the transition to chRMT by substituting the
random--matrix results for the microscopic spectral one-- and
two--point functions appearing in Eq.~(\ref{e2}).  For the quenched
chGSE, we have \cite{Naga95}
\begin{equation}
\rho_1(x)= 2x^2 \int_0^1ds\: s^2\int_0^1dt \left[ J_0(2stx)J_1(2sx)-
tJ_0(2sx)J_1(2stx)\right]
\end{equation}
and
\begin{equation}
  \tau_2(x,y)=(2xy)^2 \left[ S(x,y)S(y,x)+I(x,y)D(x,y)\right]
\end{equation}
with
\begin{eqnarray}
S(x,y)&=&\int_0^1 ds\: s^2 \int_0^1 dt \left[ J_0(2stx)J_1(2sy)
- tJ_0(2sx)J_1(2sty)\right] \\
I(x,y)&=&\int_0^1 ds\: s \int_0^1 dt \left[ J_0(2stx)J_0(2sy)
-  J_0(2sx) J_0(2sty)\right]  \\
D(x,y)&=&\int_0^1 ds\: s^3 \int_0^1 dt \:t \left[ J_1(2stx)J_1(2sy)
-  J_1(2sx) J_1(2sty)\right] \:,
\end{eqnarray}
where $J$ denotes the Bessel function.  A tedious calculation leads to
the result
\begin{eqnarray}
  \chi^{\rm disc}&=& 4u^2 \int_0^1 ds\: s^2K_0(2su) \int_0^1 dt\:
  I_0(2stu)\Bigl\{s(1-t^2)\nonumber\\
  &&\hspace*{50pt}+4K_0(2u)\left[I_0(2su)+tI_0(2stu)\right]
  -8stI_0(2stu)K_0(2su)\Bigr\}\nonumber \\
  &&-4u^2K_0^2(2u) \left[ \int_0^1 ds \:I_0(2su)\right]^2,
\label{e1}
\end{eqnarray}
where $I$ and $K$ are modified Bessel functions. To the best of our
knowledge, Eq.\,(\ref{e1}) presents a novel result. The disconnected
susceptibility is the result of strong cancellations between the two
terms in Eq.\,(\ref{e2}). For this reason, $\chi^{\rm disc}$ is 
particularly sensitive to deviations from chRMT and well-suited for the
determination of $\lambda_{\rm RMT}$. 

We turn to a comparison of $\chi^{\rm disc}$ as predicted from
Eq.\,(\ref{e1}) with lattice data. As mentioned above, this is the
first time that such a comparison has been made. In Ref.\,\cite{Jac98},
Osborn and Verbaarschot presented calculations for $\chi^{\rm disc}$
from the instanton liquid model, an effective model for QCD. Their
results show certain features which are difficult to interpret and
which may be due to finite--size effects as they suggest. 

\begin{figure}
  \centerline{\epsfig{figure=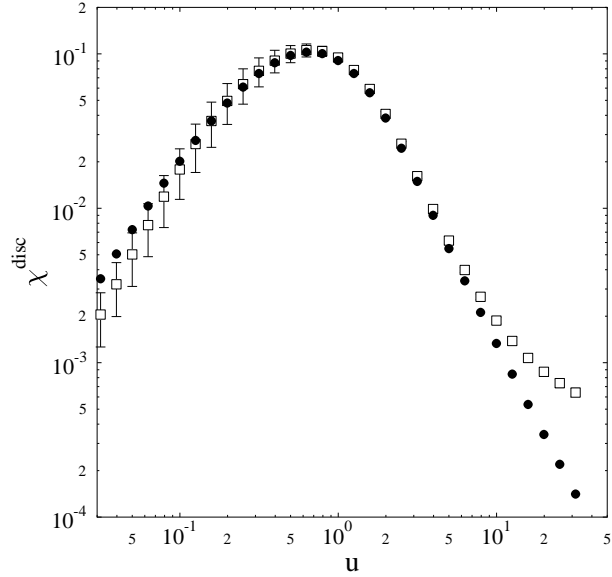,width=80mm}}
  \caption{The scaled disconnected susceptibility plotted versus the 
    scaled valence quark mass. The open squares are lattice data;
    the dots are the chRMT prediction.  The data consist of 1416 complete
    spectra on an $N=10^4$ lattice with $\beta=2.00$ and
    $\Sigma=0.1247$.}
  \label{fig1}
\end{figure}  
Figure~\ref{fig1} shows the dependence of $\chi^{\rm disc}$ on the
scaled valence quark mass $u$, defined below Eq.\,(\ref{e2}), for a
typical example, $L=10$ and $\beta=2.0$.  Here and below, the values
for $\Sigma$ are taken from Ref.~\cite{r5}.  (Note that the
eigenvalues in \cite{r5} were measured in units of $1/(2a)$.)  The
results shown in Fig.~\ref{fig1} were obtained without spectral
unfolding. We have also unfolded the lattice data, but the resulting
differences in $\chi^{\rm disc}$ are negligible since the sums in
Eq.\,(\ref{e4}) are dominated by small eigenvalues for which the
spectral density is approximately constant.  Hence, details of the
unfolding procedure are irrelevant for the present investigation.

\begin{figure}
  \centerline{\epsfig{figure=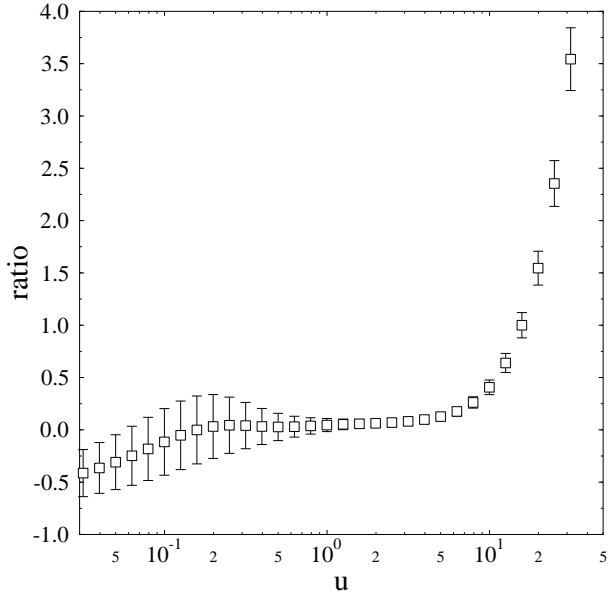,width=80mm}}
  \caption{The relative difference, Eq.\,(\ref{ratio}), of the scaled
    disconnected susceptibilities for the lattice simulation (using the 
    data from Fig.\,\ref{fig1} with $N=10^4$, $\beta=2.00$, and 
    $\Sigma=0.1247$) and chRMT.}
  \label{fig2}
\end{figure}

We note that the uncertainties in the Monte-Carlo data are correlated:
The entire set of dots in Fig.\,\ref{fig1} would shift up or down
within the range indicated by the error bars if, e.g., the lowest
eigenvalue were allowed to move within its statistical error.  Our
interest is focussed on the systematic deviations visible above
$u\approx 7$. In order to determine these deviations, we show in
Fig.~\ref{fig2} the ratio
\begin{equation}
  \label{ratio}
{\bf ratio}=\left( \chi^{\rm disc}_{\rm lattice}-\chi^{\rm disc}_{\rm RMT}
\right) / \left(  \chi^{\rm disc}_{\rm RMT} \right) \:. 
\end{equation}
Deviations of this ratio from zero determine $\lambda_{\rm
  RMT}/\Delta$.  The errors in Fig.\,\ref{fig2} are jackknife
estimates. Two features in the figure are striking. (i) Below the
lowest eigenvalue of the Monte--Carlo sample, the errors are too
small. (ii) For very small values of $u$, one observes a systematic
deviation between the lattice results and the chRMT prediction. These
features are artefacts of limited statistics and have the following
cause. The asymptotic chRMT result for very small values of $u$ is
\begin{equation}
\chi^{\rm disc}\rightarrow (2u)^2 \left[ -{1\over 3} (\ln u +
\gamma)-{1\over 12}\right]
\end{equation}
where $\gamma$ is Euler's constant. The logarithmic term is generated
by the small but finite eigenvalue density at small $u$, see
Eq.\,(\ref{e2}).  However, in a given Monte--Carlo simulation there is
always one smallest eigenvalue, $\lambda_{\rm min}$.  For values of
$u$ smaller than $\lambda_{\rm min}N\Sigma$, the logarithmic
contribution can no longer be obtained from the lattice data, see
Eq.\,(\ref{e4}).

\begin{figure}
  \centerline{\epsfig{figure=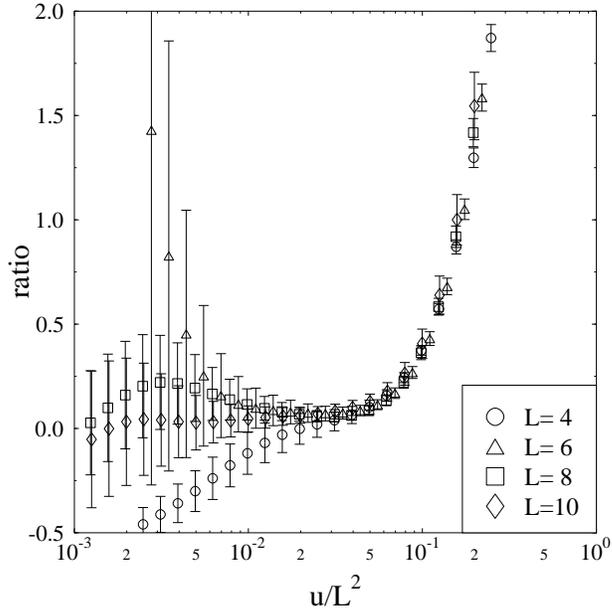,width=80mm}}
  \caption{The relative difference of the scaled disconnected
    susceptibilities plotted versus $u/L^2$ for $\beta=2.00$ and four
    different lattice sizes, $N=4^4$, $6^4$, $8^4$, and $10^4$.}
  \label{fig3}
\end{figure}

Let $u_{\rm RMT}=\pi\lambda_{\rm RMT}/\Delta$ be the value of $u$ at
which the strong deviation observed in Fig.~\ref{fig2} sets in.
According to Eq.\,(\ref{u}), $u_{\rm RMT}$ should scale with $L^2$
\cite{Zahed98,Jac98}.  To check this prediction, we have plotted in
Fig.~\ref{fig3} the ratios defined in Eq.\,(\ref{ratio}) for $L=4$, 6,
8, and 10 as a function of $u/L^2$. Obviously, all data fall on the
same curve confirming our expectation.

\begin{figure}
  \centerline{\epsfig{figure=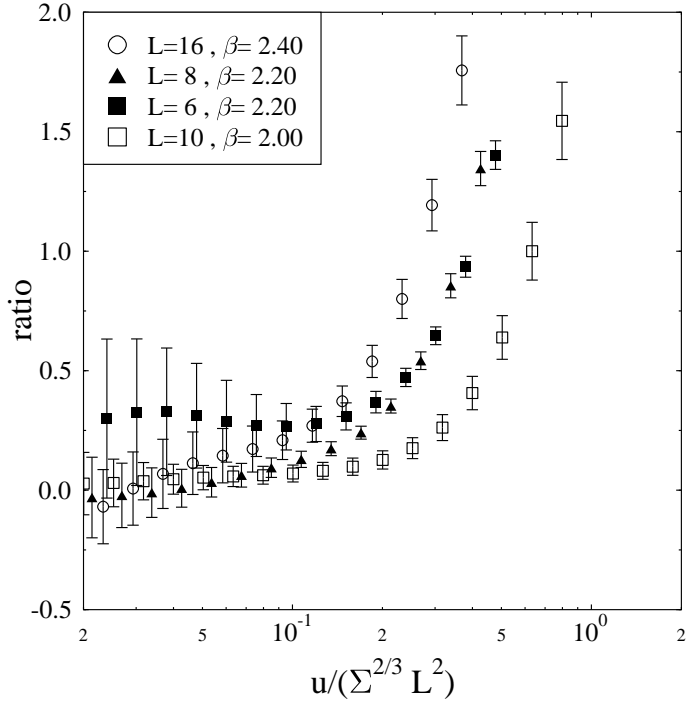,width=90mm}}
  \caption{The relative difference of the scaled disconnected
    susceptibilities plotted versus $u/(\Sigma^{2/3}L^2)$ for the data 
    of Fig.\,\ref{fig3} and additional data for $\beta=2.2$,
    $\Sigma=0.0556$ on $6^4$ and $8^4$ lattices and for $\beta=2.4$,
    $\Sigma=0.00863$ on a $16^4$ lattice.}
  \label{fig4}
\end{figure}

In order to compare results for different values of $\beta$, we note
that $u_{\rm RMT}$ is dimensionless but should be proportional to
$L^2$.  The latter quantity should scale with $a^2$, where $a$ depends
on $\beta$.  Furthermore, in the scaling regime one would expect that
$\Sigma$ scales with $a^{-3}$.  This suggests that $u_{\rm
  RMT}/(\Sigma^{2/3}L^2)$ should be independent of $\beta$ in the
scaling regime. Figure~\ref{fig4} demonstrates that this expectation
is not supported by the data.
\begin{figure}
  \centerline{\epsfig{figure=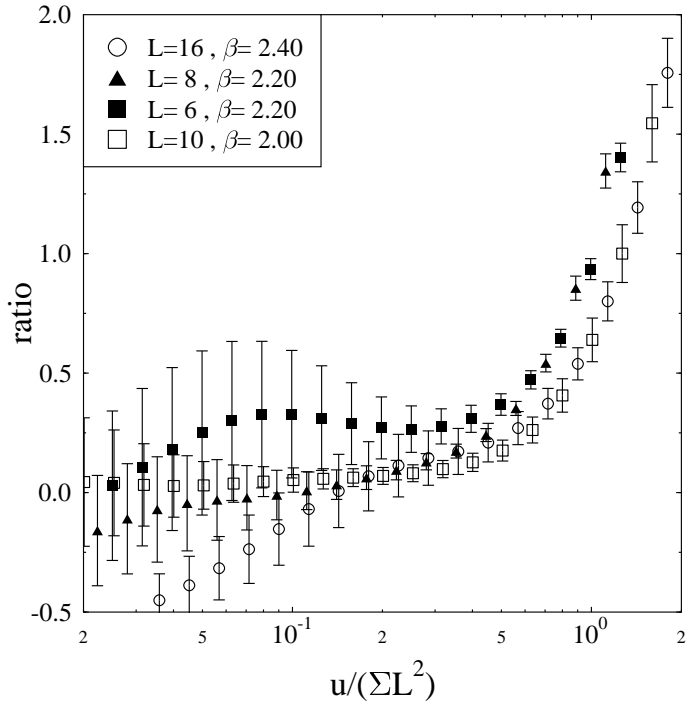,width=90mm}}
  \caption{The data of Fig.\,\ref{fig4} plotted versus $u/(\Sigma L^2)$.}
  \label{fig5}
\end{figure}
It is perhaps not too surprising that simple scaling does not work,
because the dynamics on the lattice changes in a highly complicated
manner between $\beta=2.0$ and $\beta=2.4$. The theoretical
expectation of Eq.\,(\ref{u}) is that $u_{\rm RMT}$ should scale with
$f_{\pi}^2L^2$ \cite{Zahed98,Jac98}.  A careful check of this
expectation would require the determination of $f_{\pi}$ for the
lattice sizes and $\beta$ values we have used. We have not done this.
Instead, we make use of the observation \cite{Bil} that $f_{\pi}^2$
(in lattice units) scales approximately like $\Sigma$ for the range of
$\beta$ considered here.  The plots in Fig.\,\ref{fig5} showing the
results for different $\beta$ versus $u/(\Sigma L^2)$ support this
view.

Billoire et al.~\cite{Bil} suggest that $f_{\pi}^2=\Sigma/3.4$ in
lattice units.  If one interprets Fig.\,\ref{fig5} as indicating that
$u_{\rm RMT}/(\Sigma L^2)$ is roughly 0.5, this implies (taking into
account a factor of $1/2$ from our normalization of the eigenvalues)
that
\begin{equation}
\lambda_{\rm RMT}/\Delta \approx 0.3\, f_{\pi}^2L^2 \ .
\end{equation}
This result in quenched SU(2) is in agreement with the order of
magnitude estimate $\lambda_{\rm RMT}/\Delta\sim f_{\pi}^2L^2/\pi$
from Refs.\,\cite{Zahed98,Jac98}.

\begin{figure}
  \centerline{\epsfig{figure=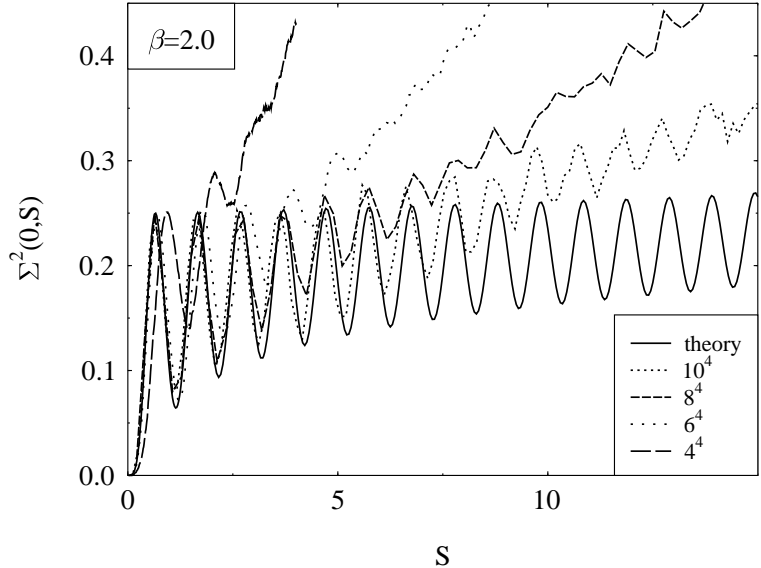,width=100mm}}
  \caption{Comparison of the number variance, $\Sigma^2(0,S)$, 
           predicted by chRMT with the results for the simulations used in 
           Fig.\,\ref{fig3}.}
  \label{fig6}
\end{figure}

\begin{figure}
  \centerline{\epsfig{figure=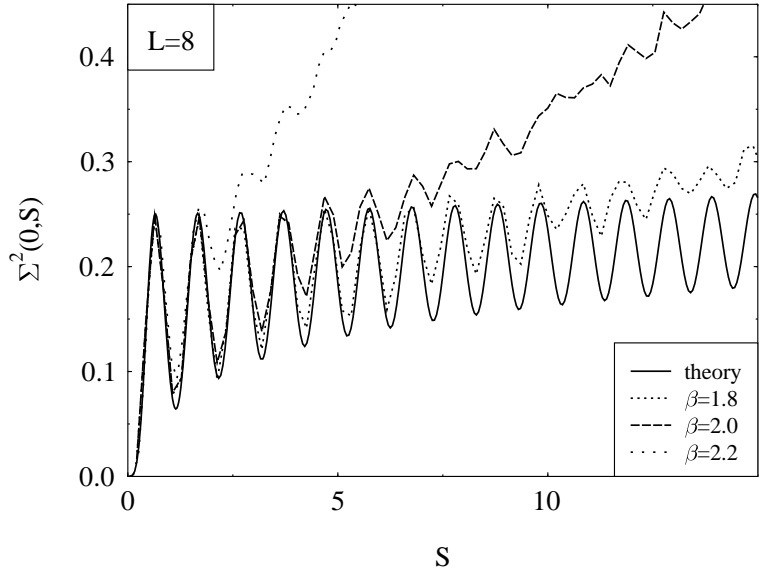,width=100mm}}
  \caption{Same as Fig.~\ref{fig6} but keeping $L$ fixed and varying
    $\beta$.} 
  \label{fig7}
\end{figure}

We now turn to the number variance which is defined as $\Sigma^2(0,S)
= \langle (N(I)- \langle N(I) \rangle)^2 \rangle$ \cite{Ma}.  Here,
$I$ is the interval $I=[0,S]$, $N(I)$ is the number of eigenvalues in
$I$, and the angular brackets denote the ensemble average.  In
contrast to $\chi^{\rm disc}$, unfolding is important for the
$\Sigma^2(0,S)$ statistic since it leads to a significant extension of
the length of the interval $I$ for which $\Sigma^2(0,S)$ can be
determined. We unfolded the spectrum by fitting the unfolding function
to the average of the spectrum over all configurations.
Figure~\ref{fig6} shows that the critical value, $S_{\rm RMT}$, for
which deviations from chRMT are observed increases with $L$.
In Fig.\,\ref{fig7} we see that $S_{\rm RMT}$ decreases with
increasing $\beta$ as expected.  
Detailed analysis of all available data sets shows that  
\begin{equation}
S_{\rm RMT} \approx (0.3-0.7)\,\Sigma L^2
\end{equation}
is consistent with the data. Hence, $S_{\rm RMT}$ and $u_{\rm RMT}$
are perfectly consistent.

The use of the $\Sigma^2(0,S)$ statistic may be conceptually more
appealing because the analogous quantity in mesoscopic systems is
directly related to the Thouless energy.  However, our analysis shows
that the susceptibility appears to be better suited for a quantitative
determination of the cross--over point from universal to
non--universal behavior.

In conclusion, we have provided the first direct determination of the
scale, $\lambda_{\rm RMT}$, which limits the validity of random matrix
descriptions of lattice QCD.  This quantity has the correct
$L^2$-scaling.  Moreover, $\lambda_{\rm RMT}$ seems to scale rough\-ly
with $f_{\pi}^2$ as expected on the basis of the
Gell-Mann--Oakes--Renner relation.

It would be very interesting to perform a detailed analysis of the
lattice data in the diffusive regime, i.e., above $\lambda_{\rm RMT}$,
to check the predictions of Ref.~\cite{Zahed98} for this regime and to
investigate possible differences between the numerical results of
Ref.~\cite{Jac98} for the instanton liquid model and the lattice
data.  Such an analysis will be the subject of future work.

It is a pleasure to thank F. Karsch and J.J.M. Verbaarschot for
stimulating discussions. This work was supported in part by DFG and
BMBF. SM, AS and TW thank the MPI f\"ur Kernphysik, Heidelberg, for
hospitality and support. The numerical simulations were performed on a
CRAY T90 at the Forschungszentrum J\"ulich and on a CRAY T3E at the
HLRS Stuttgart.

\end{document}